\documentclass[final,5p,times]{elsarticle}

\usepackage{amssymb}
\usepackage{lipsum}

\journal{Astroparticle Physics}

\begin{document}

\begin{frontmatter}



\title{Editorial\\
\emph{Cosmic Rays: High Energy Particle Interactions in the Atmosphere - Memorial Issue for Prof. Dr. Thomas K. Gaisser}}


\author[first,second]{Frank G.~Schr\"oder}
\author[first]{Serap Tilav}
\affiliation[first]{organization={Bartol Research Institute, Department of Physics and Astronomy, University of Delaware},
            city={Newark},
            state={Delaware},
            country={Unites States of America}}
\affiliation[second]{organization={Institute for Astroparticle Physics, Karlsruhe Institute of Technology (KIT)},
            city={Karlsruhe},
            country={Germany}}



\end{frontmatter}





This special issue of the \emph{Astroparticle Physics} journal is dedicated to the memory of Thomas Korff Gaisser, the Martin A. Pomerantz Professor Emeritus of Physics at the University of Delaware. 

Tom was one of the most prominent scientists in cosmic-ray and astroparticle physics, and also one of the founding editors of this very journal. A theoretical particle physicist by training, he dedicated his career to cosmic-ray physics. 
He worked on the phenomenology of high-energy particle interactions in the atmosphere, modeling extensive air showers, including the famous \lq Gaisser-Hillas\rq~function describing their longitudinal profile.
A focus of his work were muon and neutrino fluxes resulting from cosmic-ray showers in the atmosphere, and their measurement with specialized experiments at the South Pole.

Tom was born on March 12, 1940 in Evansville, Indiana. He graduated from Wabash College with a B.A.~in physics in 1962. He earned his M.Sc. degree in 1965 as a Marshall Scholar at the University of Bristol, UK. and his Ph.D. at Brown University in 1967 for a thesis on \lq Solutions of a Model Field Theoretical Equation for the Neutron-Proton Mass Difference\rq.
He was a research associate at M.I.T. in 1967-1969, and then as a NATO postdoctoral fellow at the University of Cambridge, UK, in 1969-1970.
He was hired as Assistant Professor in 1970 at the Bartol Research Foundation, which was at Swarthmore College at that time until it moved in 1977 to the University of Delaware where it was renamed to Bartol Research Institute and integrated as a center of the Department of Physics and Astronomy in 2000. 
At Bartol, Tom made the transition to cosmic-ray physics and successfully continued the legacy of its director then, Martin A.~Pomerantz, who was a pioneer of astrophysics in Antarctica.
Tom stayed at Bartol for his remaining career, was promoted to Associate Professor in 1974, to Full Professor in 1979 and became the Martin A.~Pomerantz Professor of Physics in 2001. 

Tom mastered the technique of modeling complex phenomena in elegant and transparent analytic or semi-analytic formulations. 
He is one of the fathers of the Sibyll hadronic interaction model, whose newest version remains a widely used state-of-the-art model for the simulation of air showers.
His calculations and estimates of particle fluxes helped in developing detection techniques and designing experiments, and his phenomenological models provided interpretation of complex data, needed to extract both particle and astrophysics from cosmic-ray and neutrino measurements.
Although a theorist, Tom traveled for more than 10 seasons to Antarctica and actively participated in the design and construction of experiments at the South Pole, in particular, the South Pole Air Shower Experiment (SPASE), the Antarctic Muon And Neutrino Detector Array (AMANDA), and the IceCube Neutrino Observatory with its IceTop surface array measuring cosmic-ray air showers. 
The \lq Gaisser Valley\rq, geographical location in Antarctica, is named after him for his devotion to Antarctic science.

Tom was named a Fellow of the American Physical Society in 1984 for \lq seminal contributions to our current understanding of the nature of the diverse interactions of cosmic rays with very high energies, and of their astrophysics implications\rq. He received the O’Ceallaigh Medal in 2005 and the Homi Bhabha Medal and Prize in 2015 for his distinguished contributions to Cosmic Ray Physics by the Commission on Cosmic Rays of the International Union of Pure and Applied Physics (IUPAP).  
He had a Leverhulme Visiting Professor position at University of Oxford in 2002 and was at the Humboldt University of Berlin with the Alexander von Humboldt Research Prize in 2009. 
He served the astroparticle physics community in various leadership positions, including as chair of the IUPAP C4 Commission in 1997-1999, spokesperson of the IceCube Collaboration in 2007-2011, and member of the Franklin Institute Committee on Science and the Arts. 

Tom published over 400 articles in refereed journals and conference proceedings, many of these together with his colleague Todor Stanev. 
The number of Tom's publications goes up to over 800 including IceCube Collaboration papers and proceedings. 
In 1976, he organized and edited the proceedings of an interdisciplinary meeting at Bartol \lq  Cosmic Rays And Particle Physics\rq~to foster the interaction between these fields, and in 2017 he edited a review volume with Albrecht Karle, \lq Neutrino Astronomy: Current Status, Future Prospects\rq.
Tom is renowned for his book \lq Cosmic Rays and Particle Physics\rq, a classic reference on everyone's desk in the field. The book was first published in 1990, and an updated and expanded second version came out in 2016, coauthored with Ralph Engel and Elisa Resconi. 

Thomas K.~Gaisser died on February 20, 2022 in his home in Pennsylvania at the age of 81.
He was truly an inspiring scientist. A memorial meeting was held in his honor in Newark, Delaware, in April 2023 to celebrate his scientific achievements over five decades. This special issue is a collection of five refereed articles as reflections from the meeting:

\begin{itemize}
    \item \emph{Thomas K. Gaisser, a pioneer of particle astrophysics} by Francis Halzen and Paolo Lipari, long-term collaborators of Tom, reviews his pioneering contributions to the birth and development of astroparticle physics \cite{Halzen:2024usa}.
    \item \emph{Cosmic-Ray Physics at the South Pole} by Dennis Soldin, Paul Evenson, Hermann Kolanoski, and Alan Watson reviews recent and historic cosmic-ray experiments at the South Pole and their results, as Tom Gaisser had a large impact on these experiments, in particular, AMANDA and IceCube with their surface arrays SPASE and IceTop for cosmic-ray air showers \cite{Soldin:2023lbr}.
    \item \emph{Atmospheric muons and their variation with temperature} by Stef Verpoest, Dennis Soldin, and Paolo Desiati publishes one of Tom's last research work. It builds on text from two manuscripts which he had started himself, but was not able to complete \cite{Verpoest:2024dmc}.
    \item \emph{Sibyll$^\bigstar$} by Felix Riehn, Anatoli Fedynitch, and Ralph Engel presents the latest variety of the Sibyll hadronic interaction model, where Tom was one of the original developers of this model \cite{Riehn:2024prp}.
    \item \emph{Loaded layer-cake model for cosmic ray interaction around exploding super-giant stars making black holes} by M.L. Allen, Peter L.~Biermann, and coauthors traces back to discussions between Peter Biermann, Todor Stanev, and Tom Gaisser \cite{Allen:2024qbk}.
\end{itemize}

\section*{Acknowledgements}
We would like to thank Elsevier and the team of its Astroparticle Physics journal for enabling this special issue. We also thank all authors contributing through their articles and all attendees of the memorial meeting, which was kindly supported by the University of Delaware through the College of Arts and Sciences, the Department of Physics and Astronomy, and the Bartol Research Institute.

\bibliographystyle{elsarticle-num} 
\bibliography{bibliography}






\end{document}